\documentclass[copyright,creativecommons]{eptcs}

\providecommand{\event}{M. Bujorianu and M. Fisher (Eds.):\\ \hbox{\quad}Workshop on Formal Methods for Aerospace (FMA)}
\providecommand{\volume}{20}
\providecommand{\anno}{2010}
\providecommand{\firstpage}{16}
\providecommand{\eid}{2}
\usepackage{breakurl}        

\usepackage{epsfig}

\newcommand{\mdl}[1]{#1}

\newcommand{\logscope}{{\sc LogScope}}
\newcommand{\logmaker}{{\sc LogMaker}}
\newcommand{\ruler}{{\sc Ruler}}
\newcommand{\rcat}{{\sc Rcat}}
\newcommand{\rmor}{{\sc Rmor}}

\title{An Entry Point for Formal Methods:\\ Specification and Analysis of Event Logs}

\author{
Howard Barringer
\institute{
School of Computer Science\\ 
University of Manchester\\ 
Manchester, UK}
\email{howard.barringer@manchester.ac.uk}
\and
Alex Groce
\institute{School of Electrical Engineering and Computer Science\\ 
Oregon State University\\
Corvallis, USA}
\email{alex@eecs.oregonstate.edu}
\and
Klaus Havelund \qquad\qquad Margaret Smith
\institute{Jet Propulsion Laboratory\\ 
California Institute of Technology\\
Pasadena, USA}
\email{\quad klaus.havelund@jpl.nasa.gov \quad\qquad margaret.h.smith@jpl.nasa.gov}
}

\def\titlerunning{Specification and Analysis of Event Logs}
\def\authorrunning{Barringer, Groce, Havelund \& Smith}

\begin{document}

\maketitle

\begin{abstract}
Formal specification languages have long languished, due to the grave
scalability problems faced by complete verification methods.  Runtime
verification promises to use formal specifications to automate part of the
more scalable art of testing, but has not been widely applied to real
systems, and often falters due to the cost and complexity of
instrumentation for online monitoring.  In this paper we discuss work
in progress to apply an event-based specification system to the
logging mechanism of the Mars Science Laboratory mission at JPL.  By
focusing on log analysis, we exploit the ``instrumentation'' already
implemented and required for communicating with the spacecraft.  We
argue that this work both shows a practical method for using formal
specifications in testing and opens interesting research avenues,
including a challenging specification learning problem.
\end{abstract}

\section{Introduction}

NASA's Mars Science Laboratory Mission (MSL), now scheduled to launch
in 2011 \cite{msl}, relies on a number of different mechanisms used to
command the spacecraft from Earth and to understand the behavior of
the spacecraft and rover.  The primary elements of this communication
system are commands sent from the ground, visible
events emitted by flight software (essentially
formalized {\tt printf}s in the code) \cite{exploit}, snapshots of the 
spacecraft state, and data products --- files
downlinked to earth (e.g., images of Mars or science instrument data).
All of these elements may be thought of as spacecraft
\emph{events}, with a canonical timestamp.  Testing the flight
software (beyond the unit testing done by module developers) usually
relies on observing these indicators of spacecraft behavior.  As
expected, test engineers cannot ``eyeball'' the hundreds of thousands
of events generated in even short tests of such a complex system.
Previously, test automation has relied on ad hoc methods ---
hand-coded Python \cite{python} scripts using a framework to query the ground
communications system for various kinds of events, as a test proceeds.
This was the state-of-the-art at the time members of our group, the
Laboratory for Reliable Software, joined the MSL team, as developers
and test infrastructure experts.  Lengthy collaboration with test
engineers convinced us that something better was required: the test
script approach required large amounts of effort, much of it
duplicated, and often failed due to changes in event timing and
limitations of the query framework.  Moreover, the scripts, combining
test input activity, telemetry gathering, and test evaluation, proved
difficult to read --- for other test engineers and for developers and
systems engineers trying to extract and understand (and perhaps fix) a
specification.  Runtime verification using formal specifications
offered a solution, and the MSL ground communications system suggested
that we exploit an under-used approach to runtime verification:
offline examination of event logs already produced by system
execution.  In the remainder of this paper we report on two aspects of
this project --- in Section \ref{reqs} we discuss the general idea of
event sequences as requirements, and our specification methodology;
Section \ref{framework} gives a brief introduction to our application
of these general ideas to MSL and our new specification language.

\section{Event Sequences as Requirements}
\label{reqs}

Systems verification consists of proving that an artifact (hardware
and/or software) satisfies a specification.  In mathematical terms we
have a model $M \in \mdl{ML}$ (for example the complete system) in
some model language $\mdl{ML}$, and a specification $S \in \mdl{SL}$
in some specification language $\mdl{SL}$, and we want to show that
the pair $(M,S)$ is member of the satisfaction relation 
$\models\; \subseteq\; \mdl{ML} \times \mdl{SL}$, also typically written: $M
\models S$. The general problem of demonstrating correctness of a combined
hardware/software system is very hard, as is well-known. Advanced
techniques such as theorem proving or model checking tend not to scale
for practical purposes. Extracting abstract models of the system, and
proving these correct, has been shown to sometimes be useful, but also
faces scalability problems when dealing with real systems or complex
properties.  The problem is inherently difficult because the models
are complex -- because the behavior of systems is complex.  The problems of full verification have long limited the adoption of formal specification.

In \emph{runtime verification} a specification is used to analyze a
single execution of a system, not the complete behavior.  In this case
a model is a single execution trace $\sigma$, and the verification problem
consists of demonstrating that $\sigma \models S$, a much simpler problem
with scalable solutions. The original model $M$ (the full system) can
be considered as denoting the set of all possible runs $\sigma$, of which
we now only verify a subset. This approach of course is less
ambitious, but seems to be a practical and useful deployment of formal
specification.  Though these observations are well known, they are
less often taken into practice. It is still rare to observe the
application of formal specification --- even for runtime
verification. There may be several reasons for this, one of which is
the problem of program instrumentation. A large body of research
considers the problem: how to we produce traces to verify? There is,
however, a much simpler approach, namely to use logging information
that is already generated by almost any computer system when it is
tested. Our {\em first recommendation} is therefore that runtime
verification should be used to formally check logged data.  Our {\em
second recommendation} is that logging and requirements engineering
should be connected in the sense that requirements should be testable
through runtime verification of logs. The common element is the {\em
event}: requirements should be expressed as predicates on sequences of
events, and logging should produce such sequences of events, thereby
making the requirements testable. This means that logs preferably
should consist of events with a formal template, connected to the
original requirements. Note, however, that it is possible to extract
formalized events from chaotic logging information produced by the
usual ad hoc methods --- using, e.g., regular expressions.

The obvious scientific and methodological question is: what kind of
information should a log contain, in order to help verify
requirements?  We shall adopt the simple view that a log is a sequence
of events, where an event is a mapping from names to values of various
types (integers, strings, etc.):

\[
   \begin{array}{rcl} 
      Log & = & Event^* \\
      Event & = & Name \rightarrow Value \\
   \end{array}
\]

\begin{figure}[htb]
\begin{center}
\includegraphics[width=0.70\textwidth]{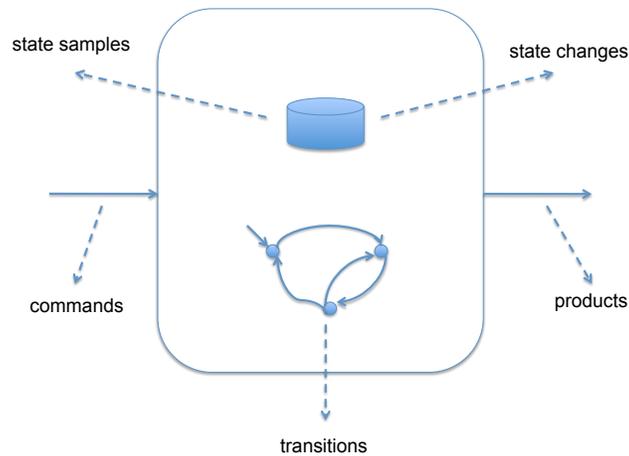}
\end{center}
\caption{observable events of a system}
\label{fig:observable-events}
\end{figure}

\noindent This definition is quite general. An event can carry information
about various aspects of the observed system. Events can be classified
into various kinds by letting a designated name, i.e. {\em kind}, map
to the kind. In the context of MSL, five forms of events are used, see
Figure \ref{fig:observable-events}. We claim that these five event
kinds are generally applicable to any system being monitored. The five
forms of events are: (1) commands (input) issued to the monitored
system, (2) products (output) delivered by the system, (3) periodic
samplings of the state of the system, such as the value of
continuous-valued sensors (like position coordinates), (4) changes to
the state of the system, those changes that are observable, and (5)
transitions performed by the system (also referred to as EVent Reports - EVRs), 
for example when {\tt printf} 
statements would normally be used to record an important event. 
The two forms of observation of the state (3, 4) could
potentially be regarded as one kind of observation: that of the state
of the system at any point in time. The state observations (3, 4) and
the transitions (5) are internal events, while the commands (1) and
products (2) are external events.

We have developed a specification language, and corresponding
monitoring system, to be presented in the next section, which have
been applied by testing engineers within the MSL
project. The specification language consists of a mixture of automata
and temporal logic with elements of regular expressions. The logic can
specifically refer to the data in events. This mixture seems to be
attractive for the engineers. In the longer term, we argue that such a
specification language and monitoring system should be used in
combination with a systematic logging discipline, such that
\emph{requirements are formulated in terms of events}, and, minimally, \emph{these events}
should be produced as part of logging. These observations may not
appear to be ground-breaking, and to some extent have the flavor of
{\em ``yes of course''} --- but as it turns out, considering current
practices in software projects, they may be rather ground-breaking. A
successful formal verification story always relies on finding the
proper mixture of formality and common practice, as well as the right
specification language for the task. We believe that the work
described in this short paper has shown the potential for 
being such a success story.


\section{Framework}
\label{framework}

MSL's ground software stores all events in a SQL database, which we
interpret as a chronologically ordered sequence of events --- a log.
Our Python framework, called \logscope{}, allows us to check logs for
conformance to a specification and to ``learn'' patterns from logs.
The architecture of \logscope{} divides functionality into a
\logmaker{} tool, specific to MSL, and a core \logscope{} module for
checking logs and learning specifications, which may be applied to any
ordered event sequence.

\subsection{LogMaker}

\logmaker{} communicates with MSL's SQL-based ground software
to generate a list of events, where each {\em event} is a record
mapping field names to their values.  A special field indicates the type of the event: 
command, transition (EVR), state sampling, state change, or data product.  Note how the MSL events map to the
generalized idea of a monitored system shown in Figure
\ref{fig:observable-events}.  The log extractor sorts events according
to spacecraft event times, since the order in which events are
received by ground communications software does not correspond to the
order in which events are generated on-board (due to varying
communication priorities).  Further analysis annotates the log with
meta-events for ease of use in specification, and uses spacecraft
telemetry to assign a spacecraft time to ground events.  We hope to
extend and exploit previous work on monitoring distributed systems
with multiple clocks \cite{Sen06} to influence flight software's use
of telemetry to ensure that effective event ordering is always
possible.

\subsection{Monitoring}

\begin{figure}
{\small
\begin{verbatim}
  pattern CommandSuccess:
    COMMAND{Type : "FlightSoftwareCommand", Stem : x, Number : y} => 
       {
          EVR{Dispatch : x, Number : y}, 
          [ 
             EVR{Success : x, Number : y},
             not EVR{Success : x, Number : y}
          ],
          not EVR{DispatchFailure : x, Number : y},
          not EVR{Failure : x, Number : y}
       }
\end{verbatim}
}
\caption{A generic specification for flight software commands.}
\label{command}
\end{figure}

The monitoring system of \logscope{} takes two arguments: (1) a log
generated by {\em logmaker}, and (2) a specification.  Our
specification language supplies an expressive rule-based language,
which includes support for state machines, and a higher-level (but
less expressive) {\em pattern language}, which is translated into the
more expressive rule-based language before monitoring.

Specifications in the pattern language are easy for test engineers and
software developers to read and write.  Figure \ref{command} illustrates
a pattern. The {\tt CommandSuccess} pattern requires that following every command
event (meaning a command is issued to the flight software),
where the {\tt Type} field has the value {\tt
"FlightSoftwareCommand"}, the {\tt Stem} field (the name of the
command) has a value {\tt x} ({\tt x} will be \emph{bound} to that
value), and the {\tt Number} field has a value {\tt y} (also a binding
variable), we must see ({\tt =>}) --- in any order, as indicated
by set brackets \verb+{...}+ --- (1) a dispatch of command {\tt x}
with the number {\tt y}; (2) a success of {\tt x}/{\tt y}, and {\em after that} no more
successes --- the square brackets \verb+[...]+ indicate an ordering of
the event constraints. Furthermore, (3) we do not want to see any
dispatch failures for the command; and finally (4) we do not want to
see any failures for the command.

Interesting features of the language include its mixture (and nesting)
of ordered and unordered event sequences of event constraints,
including negations, and its support for testing and capturing data
values embedded in events.  The pattern language is translated into our
rule-based language derived from the \ruler{} specification language
\cite{barringer-ruler-07,barringer-ruler-journal-08,BarringerFM09}.  A subset
of this language defines state machines with parameterized events and
states, where a transition may enter many target states ---
essentially alternating automata with data.  The language is also
inspired by earlier state-machine oriented specification/monitoring
languages, such as \rcat{} \cite{smith-havelund-rcat-08} and
\rmor{} \cite{havelund-rmor-08}. 


In addition to exact requirements on field values, our language
supports user-defined predicates (written in Python) that may take
field values and bound variables as arguments, providing very high
expressive power.
Specifications are visualized with Graphviz \cite{graphviz}, and extensive error trace reporting with
references to the log files ensures easy interpretation of detected specification
violations.

\subsection{Learning}

\logscope{} was well-received by test engineers, and was integrated
into MSL flight software testing for two modules shortly after its release.  
One important result of early use was to
alert us to the burden of writing patterns more specific than the kind
of generic rule shown above.  In order to ease this burden we
introduced a facility for \emph{learning} specifications from runs. 
Consider a test engineer or developer who runs a flight software test
one or more times. If these runs have been ``{\em good}'' runs he/she
can ``endorse'' (perhaps after making manual modifications) the
specification, and it can then be used to monitor subsequent executions.
Learning requires a notion of event equality, and users can define
which fields should be compared for testing event equality (e.g.,
exact timing is usually expected to change with new releases and
perhaps even new test executions).  We have implemented and applied a
{\em concrete learner} which learns the set of all execution sequences
seen so far (essentially a ``diff'' tool for logs).  We also expect to
learn mappings from commands to events expected in all execution
contexts --- a pattern based approach, like that of Perracotta
\cite{Perracotta}.  More ambitiously, we hope to incorporate classic
automata-learning results \cite{angluin-87} in order to generalize
specifications.

\section{Conclusions and Future Work}

The MSL ground control and observation software demonstrates an
important concept: many critical systems already implement very
powerful logging systems that can be used as a basis for 
automated evaluation of log files against requirements. Such log files
can be analyzed with scripts (programs) written using a scripting (programming) language. 
However, there seems to be advantages to using a formal specification
language, as demonstrated with this work. A systematic study could be needed,
investigating to what extent a domain specific language really is required to achieve the added benefit,
or whether a well designed Python API (or API in any other programming language)
would yield the same benefits.

\subsubsection*{Acknowledgements} 
Part of the research described in this publication was carried out at the Jet Propulsion Laboratory, California Institute of Technology, under a contract with the National Aeronautics and Space Administration.

\vspace{0.5cm}

\noindent
Thanks are due to many members of the Mars Science Laboratory Flight Software team, including 
Chris Delp, 
Gerard Holzmann,
Rajeev Joshi,
Cin-Young Lee, 
Alex Moncada, 
Cindy Oda, 
Glenn Reeves,
Margaret Smith, 
Lisa Tatge, 
Hui Ying Wen, 
Jesse Wright, and
Hyejung Yun.


\bibliographystyle{eptcs} 
\bibliography{biblio}

\end{document}